\documentclass[aps,pre, preprint]{revtex4}

\usepackage{graphicx}
\usepackage{listings}
\usepackage{color} 
\definecolor{mygreen}{RGB}{28,172,0} 
\definecolor{mylilas}{RGB}{170,55,241}

\usepackage{amssymb,amsfonts,amsmath,color}

\begin{document}


\title{The time-identity tradeoff}
\author{ Nadav M.\ Shnerb}
\vspace{1.0cm}
\affiliation{Department of Physics, Bar-Ilan University, Ramat-Gan IL52900, Israel}
\begin{abstract}
Distinguishability plays a major role in quantum and statistical physics. When particles are identical their wave function must be either symmetric or antisymmetric under permutations and the number of microscopic states, which determines entropy, is counted up to permutations. When the particles are distinguishable, wavefunctions have no symmetry and each permutation is a different microstate. This binary and discontinuous classification raises a few questions: one may wonder what happens if particles are almost identical, or when the property that distinguishes between them is irrelevant to the physical interactions in a given system. Here I sketch a general answer to these questions. For any pair of non-identical particles there is a timescale, $\tau_d$, required for a measurement to resolve the differences between them. Below $\tau_d$, particles seem identical, above it - different, and the uncertainty principle provides a lower bound for $\tau_d$. Thermal systems admit a conjugate temperature scale, $T_d$. Above this temperature the system appears to equilibrate before it resolves the differences between particles, below this temperature the system  identifies these differences before equilibration. As the physical differences between particles decline towards zero, $\tau_d \to \infty$ and $T_d \to 0$.
\end{abstract}
\maketitle

\section{Introduction}

Physicists believe that nature does not make jumps. In any realistic physical system all changes must take place gradually; discontinuous jumps, like the change of density or specific heat during a phase transition, occur only in the abstract, in infinite systems.  This simple insight appears to be inconsistent with the distinction between identical and non-identical particles.

Identity is a binary feature: two objects are either identical or not. Any tiny  difference in any physical property - mass, charge, size - is enough to make two particles non-identical. Therefore,  nature \emph{does} make jumps: two particles, one with physical property ${\cal O}$ and another with ${\cal O} + \delta  {\cal O}$  are not identical particles for every finite value of $\delta {\cal O}$ and become identical, abruptly, at  $\delta  {\cal O}=0$. When other physical properties (e.g., entropy) are governed by indistinguishability of particles, these properties and the related observables also undergo a discontinuous jump at $\delta  {\cal O}=0$.

The identity of different particles manifests itself in thermal physics and quantum mechanics. Thermodynamics and statistical mechanics require the entropy to be an extensive property, and this holds only if all $N!$ permutations of $N$ identical particles are considered as a single microscopic state (\cite{gibbs1902elementary}, Ch. 15). If these particles differ from each other, even by tiny differences (ortho-and para-hydrogen, molecules with different chirality, different isotopes and so on) that may be irrelevant to the experimental setup then formally the number of possible microscopic states has to be multiplied by $N!$ and the entropy increases tremendously. The famous Gibbs paradox associated with this counting problem has many formulations~\cite{murashita2017gibbs}, but at least some authors consider this discontinuous dependence of entropy on identity as the paradox itself~\cite{van1984gibbs,denbigh1989gibbs,jaynes1992gibbs,allahverdyan2006explanation,dieks2011gibbs}.

Quantum mechanics requires the wavefunction of two identical particles, $\psi(x_1,x_2,t)$, to be either symmetric (bosons) or antisymmetric (fermions) under permutation of particles. Again, the laws of quantum physics allow for a two-particle wavefunction with arbitrary symmetry properties as long as $\delta  {\cal O} \neq 0$ (hypothetically, even if their couplings to the bosons that carry the weak force differ slightly), but once the particles attain the zero difference point their wavefunction becomes either symmetric or antisymmetric. Does nature make jumps?

Here I would like to argue against the instantaneous jump idea by pointing out the importance of measurements and their associated time scales. To distinguish between non-identical particles one has to preform a measurement, and if the differences are very small - a precise measurement. When the particles are almost-identical, as we shall show, the timescale associated with the required measurement diverges. For any physical property ${\cal O}$ (mass, charge etc.) and for any system of particles that differ from each other by $\delta {\cal O}$, there is a conjugate timescale $\tau_d$ below which an observer cannot realize that it is not a system of identical particles. For truly identical particles  $\tau_d \to \infty$, so no experiment can identify the (non-existing) differences between them. In thermal systems there is an associated temperature scale $T_d$. Above $T_d$ the system appears to equilibrate before it resolves the differences between particles, below this temperature the differences between particles manifest themselves in the out-of-equilibrium dynamics.

This paper is organized as follows. In the next section I shall discuss a few thermodynamic scenarios in which a semi-permeable membrane plays an important role. In these cases, the relevant measurements occur at the membrane that must distinguish between different particles, and the identification of the diverging timescale is easy. Section \ref{sec3} is focused on the time required for a measurement that distinguishes between two particles and stresses the restrictions imposed by the quantum uncertainty principle. Section \ref{sec4} returns to thermodynamics from a broader perspective, which allows us to identify the relevant temperature scale. Finally, the discussion section provides a summary and explains the relationships between this and former works.

\section{Distinguishably and time: the case of a semi-permeable membrane}

As pointed out by Gibbs~\cite{gibbs1878equilibrium,denbigh1989gibbs}, when two ideal and homogenous gases ($A$ and $B$) are in thermal equilibrium, their mixing may lead to an increase in the total entropy of the system. If the molecules of $A$ differ from the molecules of $B$, this increase, $\Delta S$, is larger than zero and is independent of $\delta  {\cal O}$.  On the other hand, at $\delta  {\cal O}=0$ particles are identical and $\Delta S$ suddenly collapses to zero.

This section deals with a simple example, the mixing of two gases through a semi-permeable membrane. The performances of such a membrane depend on its ability to distinguish between $A$ and $B$ particles, and the examples below illustrate the relationship between particles similarity and the time requires for their mixing.

\subsection{An irreversible process: free (Joule) expansion}

As a first example let us consider a classical thermodynamic setup. Two ideal gases $A$ and $B$, for which $N_A= N_B = N/2$, fill the two sides of a container whose total volume is $V$. A partition divides the volume into two equal parts, $V_A = V_B = V/2$. The system's temperature is fixed at $T$, so the pressures are equal, $P_A = P_B = N k_B T/V$, where $k_B$ is the Boltzmann constant.

Assume the only feature that distinguishes between $A$ and $B$ is their size. $A$ is a gas of spheres of radius $r_A$, where $B$ particles have a radius $r_B >r_A$, such that $\delta {\cal O} \equiv r_B -r_A = \delta r$.  Unless $\delta r=0$, $A$ and $B$ are distinguishable, and the entropy increases by $\Delta S =  N k_B \ln 2$ when the partition opened up. In  that case, $\Delta S$ is independent of $\delta r$. When $\delta r=0$, $A$ and $B$ particles are indistinguishable and an appropriate $1/N!$ factor must be introduced when calculating the number of microscopic states, thus entropy does not change ($\Delta S = 0$) when the partition is removed.

\begin{figure}[h]
	\centering{
		\includegraphics[width=9cm]{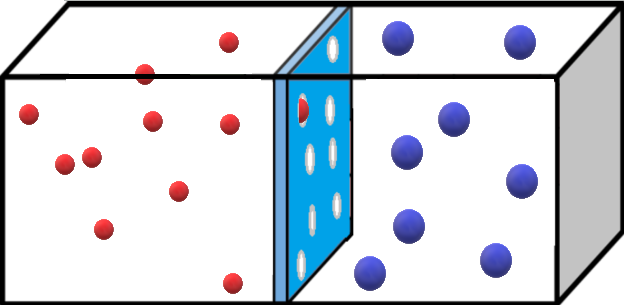}}
	\caption{\textbf{The time-identity tradeoff}: two gases, $A$ and $B$, fill the two sides of a container. $r_A$, the radius of the $A$ particles (red), is smaller than $r_B$, the radius of the $B$ particles (blue).  At $t=0$ a partition separating the two sides is replaced by a semi-permeable membrane with holes of radius $r_h$ such that $r_A < r_h <r_B$, so $A$ particles may enter the right chamber but $B$ particles cannot enter the left chamber. This procedure allows the entropy of the $A$ gas to increase and yields a net flow of particles (wind) in the right side of the chamber, from which one may extract work. The time required for an $A$ particle to slip through the membrane diverges as $r_h-r_A$ approaches zero. Therefore, as $r_B \to r_A$ the mixing process slows down and the time needed for appreciable entropy changes diverges.       \label{fig1}}
\end{figure}

How the distinction between $\delta r = 0$ and $\delta r \neq 0$ manifests itself dynamically? To understand that, let us replace the partition by a membrane permeable to $A$ particles, but not to $B$ particles (Figure \ref{fig1}).  Because ideal gas particles at the same temperature are (in effect) noninteracting and the membrane
is impermeable to $B$, the expansion of $A$ occurs as if $B$ were not present. When the system equilibrates again its entropy grows by a factor $\Delta S = N_A k_B \ln 2$.  As mentioned above, $\Delta S$ is independent of $\delta r$ as long as $\delta r >0$.

Clearly, such an irreversible expansion yields a density wave that propagates from left to right, so one may measure the associated ``wind" or even translate some of its kinetic energy to work by introducing a little wind turbine in the right side of the chamber described in Figure \ref{fig1}.

These considerations point toward an important observation: the thermodynamic distinction between identical and non-identical particles has to do with the ability of a physical measurement (here: at the membrane) to tell them apart. As long as $\delta r >0$, a semi-permeable membrane is feasible.  When $\delta r=0$ (particles are identical) no physical membrane is permeable to $A$ but impermeable to $B$.

We can equip the membrane with a little demon that opens a gate when a particle arrives from the left and closes it when it comes from the right, but this will be yet another Maxwell's demon for which the act of acquiring information requires memory which must finally be erased, a process that increases the entropy of a system~\cite{bennett1982thermodynamics,bennett1985fundamental,mandal2012work}. The membrane shown in Figure \ref{fig1} is ``passive", since it does not need to change its own internal state to allow particles to pass. Exactly because of that it becomes ineffective when $\delta r = 0$, even if $A$ and $B$ particles still have different colors or other physical properties that are irrelevant to their size.

Now we can understand what happens as $\delta r$ decreases \emph{towards} zero. Although the total change of $\Delta S$ is independent of $\delta r$, as $\delta r$ decreases the size of the perforations in the membrane's mesh must approach the size of the $A$ particles, otherwise it will allow the $B$ particles to penetrate $V_A$ as well. As the size of these holes decreases towards $r_A$, the (per collision) chance of a given $A$ particle to slip through the membrane decreases, so the time to equilibration increases. As suggested above, for each finite $\delta r$ there exist a timescale $\tau_d$ below which $\Delta S$ cannot be measured, so when $t < \tau_d$  identical and almost-identical particles yield the same physical outcomes in an experiment. Only above $\tau_d$ does the distinguishability between gas particles manifest itself in an experiment, and this timescale diverges as $\delta r \to 0$.

Note that the permeable membrane must be sensitive to the property that makes $A$ and $B$ different (e.g., radius). Membrane with other permeability criteria will function either as a closed partition (if it is impermeable to both) or as a widely open gate between the two sides. In both cases there are no density gradients and no wind to extract work from.

\subsection{A reversible process: Carnot cycle}

The semi-permeable membrane of Figure \ref{fig1} may facilitate the operation of a heat engine. Let us take, as our initial conditions, the situation after the equilibration of the $A$ gas, when the number of $A$ particles in each side of the partition is equal but the $B$ particles are still trapped in the right side of the container. Clearly the net pressure of the $A$ gas on the membrane, $P_A$, is zero. Moreover, even if the semi-permeable membrane moves, as long as the $A$ gas is at equilibrium (so the ratio between the number of $A$ particles to the left and to the right of the membrane is equal the volume ratio $V_A/V_B$) $P_A$ remains zero.

For the $B$ gas the membrane is impermeable, so it can play the role of a piston in a standard heat engine. In particular, one may implement the standard sequence of isothermal expansion (the membrane moves to the left), adiabatic expansion, isothermal compression and adiabatic compression to extract heat from a hot reservoir, put less heat in a cold reservoir and translate the difference into work.

Importantly, the efficiency of such a hypothetic heat engine (when it operates reversibly) is independent of $\delta r$ as long as this quantity differs from zero. When $\delta r=0$ the membrane becomes a perfect partition, so once it moves to the left (during an isothermal expansion, say) $P_A$ increases in the direction that opposes the movement of the membrane. Therefore, the total work extracted from a single cycle is a discontinuous function of  $\delta r$.

Again, a decrease in $\delta r$ manifest itself dynamically. While the work extracted in a single cycle is $\delta r$ independent, the time required for such a cycle diverges as $\delta r \to 0$.  To maintain $P_A=0$ as the membrane moves, and to keep all particles at the same temperature during adiabatic expansion and compression, $A$-particles must cross the membrane. When $\delta r \to 0$ this becomes the rate-determining step, thus the ability of the engine to approach its reversible (Carnot) limit is governed by $\tau_d$.

\section{Uncertainty, quantum measurement time and identical particles} \label{sec3}

The discrimination mechanism illustrated in Figure \ref{fig1} (hole size) is, of course, only an example. Many possible mechanisms may allow to distinguish between $A$ and $B$ particles that admit different physical properties, and each of these mechanisms has its own $\tau_d$. In this section I would like to claim that for each \emph{finite} detector, which implements finite fields, quantum mechanics sets a strict lower bound on $\tau_d$ given the physical differences  $\delta {\cal O}$, through one of its fundamental relationships, the time-energy uncertainty principle  $\Delta E \, \delta t \ge \hbar/2$~\cite{mandelstam1991uncertainty,cohen2006quantum}.

\begin{figure}[h]
	\centering{
		\includegraphics[width=9cm]{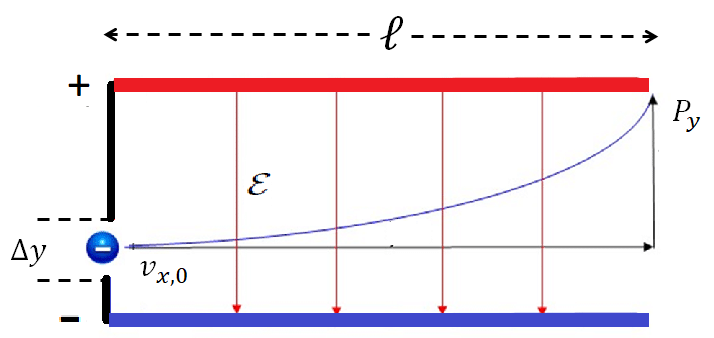}}
	\caption{\textbf{Quantum uncertainty and measurement time:} a particle of negative charge $q$ and velocity $v_{0,x}$ enters, through a slit of width $\Delta y$, a region between the two plates of a capacitor where a constant electric field ${\cal E}$ is pointing in the negative $y$ direction, so $\vec{F} = q {\cal E} {\hat y}$. The time this particle spends in the capacitor (the measurement time $t_{d,qm}$) is $ \Delta t = \ell/v_{0,x}$, where $\ell$ is the length of the plates.  The uncertainty in initial momentum in the $y$ direction is governed by the width of the slit, $\Delta P_y  \approx \hbar/\Delta y$. Therefore, to identify a charge difference  $\delta q$ the contribution of the electric field to the $y-$momentum at the exit, $P_y = \delta q \, {\cal E} \, \Delta t$, must be larger than  $\Delta P_y$. Accordingly, the measurement time must satisfy $\Delta t >\hbar/\Delta E(\delta q)$, where $\Delta E(\delta q) = \delta q \, {\cal E} \, \Delta y$ is the energy difference between the upper edge and the lower edge of the entrance slit. For any finite $\delta q$ the detection time decreases like $1/\Delta y$, but the finiteness of the detector and of the electric field sets an upper bound on $\Delta y$ and dictates a minimal value $t_{d,qm}$ for any $\delta q$.                 \label{fig3}}
\end{figure}

To measure a property (e.g., electric charge $q$) one must apply a field that couples to this charge (electric field ${\cal E}$ in that case) and track the changes in momentum due to the applied force. Therefore, for any $\delta q$ there is a corresponding energy scale $\Delta E(\delta q) = {\cal E} \, \delta q \, \Delta x$, where $\Delta x$ is the quantum uncertainty in the position of the particle. As a result (see example in Figure \ref{fig3}), the minimal time required for a measurement that resolves $\delta q$ differences between particles diverges like $\hbar/\Delta E(\delta q)$ as $\delta q \to 0$. The same is true for (almost) any other physical observable ${\cal O}$: to measure differences one must couple the observable to a relevant field, this yields an energy difference scale (force times the position uncertainty $\Delta x$) that vanishes as ${\delta \cal O} \to 0$, and the quantum measurement time diverges accordingly.

An exception to this picture, but not to the uncertainty principle, are mass measurements. Mass differences involve an inherent energy scale, $\Delta E(\delta m) = \delta m c^2$, that in principle has nothing to do with the applied gravitational field. However, as noted by Bohr in his famous argument with Einstein during the Solvay conference of 1930, the strength of the gravitational field affects the flow of time~\cite{jammer1974philosophy,aharonov2008quantum}.  As a result (see Appendix \ref{apa}), location uncertainty $\Delta x$ is translated directly to $\Delta t$  in a way that preserves the uncertainty principle. In general, thus, any measurement of $\delta {\cal O}$, for any physical observable ${\cal O}$, requires a minimal quantum measurement time,
\begin{equation}
\tau_d \geq t_{d,qm} \approx \frac{\hbar}{\delta E(\delta {\cal O})}.
\end{equation}

This brings us back to the quantum mechanical version of the question, namely the discontinuous jump in the properties of a quantum system at $\delta {\cal O} =0$. Although the wavefunction of two non-identical particles does not have any particular symmetry, to discover that $|\psi(x_1,x_2,t)|^2 \neq |\psi(x_2,x_1,t)|^2$ an observer must measure at least one relevant observable by which these two particles differ from each other. As long as $t < t_{d,qm}$ an experiment cannot \emph{discover} the violation of the required symmetry, as the experiment cannot identify the physical properties that distinguish between the particles. Only above $t_{d,qm}$ the distinction between particles emerges, and for any finite detector (finite size, finite fields) $t_{d,qm}$ increases without limit as the differences between particles vanishes.

\section{Gibbs paradox, measurement time and quantum mechanics} \label{sec4}

The insights gathered through the last two sections allow to understand better the relationships between the concept of identical particles, measurement time, the symmetries required from quantum mechanical wavefunctions and the permutation counting problem in statistical physics.

\subsection{Quantum and classical detectors} 

As the similarity between two particles increases, the time requires to distinguish between them, $\tau_d$, increases as well. The quantum mechanical uncertainty principle imposes a lower bound on $\tau_d$: given a finite detector, in which the fields are bounded from above, $\tau_d \geq t_{d,qm}$. On the other hand, in classical systems there is no such lower bound.  For each $\delta {\cal O}$, one may build an accurate finite detector with an arbitrary short $\tau_d$. Therefore, in a classical world the semi-permeable partition of Fig. \ref{fig1} may be replaced by a more sophisticated apparatus that will make the filtering time much shorter than all other timescales in the problem. In a quantum world the physical dimensions and the fields at the partition impose an upper limit to the rate at which $A$ particles percolate to the right side of the container.

In what follows I will argue that the similarity between particles manifests itself in the dynamics when $\tau_d$ becomes the rate-determining step in the equilibration process. In a given classical system we can always improve our detectors to make $\tau_d$ shorter, so the dynamical properties of a classical system with ideal detectors undergo an abrupt jump at $\delta {\cal O} =0$. Quantum mechanics prohibits the existence of such ideal detectors so the transition is necessarily gradual.

\subsection{Detection time and equilibration}  

To begin, let us define a pseudo-equilibrium state as the thermodynamic state that the system would have reached had the particles been identical. The time required to reach that state is $t_{peq}$, whereas the time to reach the true equilibrium state is $t_{eq}$. Both timescales depend, in general, on the initial conditions and on many other physical factors. $t_{peq}$ may differ from $t_{eq}$ even if $\tau_d \to 0$. For example, in the situation described in Figure~\ref{fig1} $t_{peq}$ is zero, since the partition plays no role if $A$ and $B$ are identical. Still, time is required to reach a true equilibrium state: even if $A$ particles can cross the semi-permeable partition instantaneously, they still have to spread all over the right chamber.

The interactions due to which $t_{eq}$ differs from $t_{peq}$ must involve physical processes that distinguish between particles. Therefore, as $\delta {\cal O} \to 0$, $\tau_d \to \infty$ becomes the rate-determining step in the (true) equilibration process. This may lead to a separation of timescales between $t_{eq}$ and $t_{peq}$; in the interim period, $t_{peq} \ll t \ll t_{eq}$,  the thermodynamics of a system of non-identical particles appears to be the same as the thermodynamics of a system of identical particles.

In our example (Figure~\ref{fig1}) the equilibration process involves two time-scales, the membrane detection time $\tau_d$ and the invasion time $t_{inv}$. The invasion time is the period required for the $A$ particles that already crossed the membrane to reach the right side of the container and is determined by the ratio between the linear size of the chamber and the sound velocity. For  gases the velocity of sound is approximately the typical velocity of a molecule, so $t_{inv} \propto \sqrt{m \ell^2/k_B T}$ where $\ell = V^{1/3}$ is the linear size of the system, $m$ is the mass of the particles and $k_B$ the Boltzmann constant.

\begin{figure}[h]
	\centering{
		\includegraphics[width=7cm]{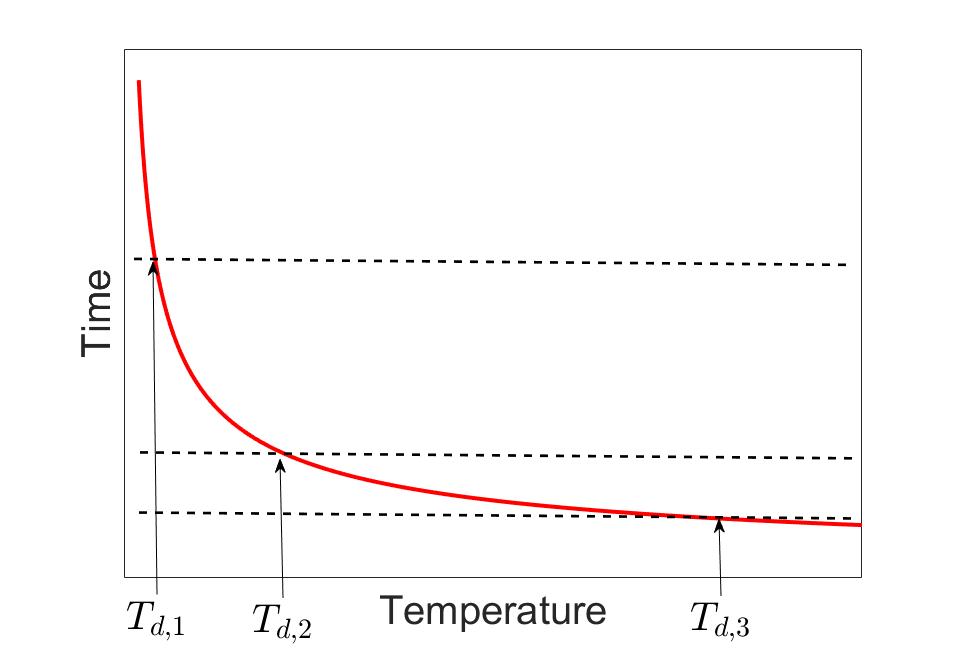}}
	\caption{\textbf{Time and temperature scales} In the example sketched in Figure \ref{fig1}, one may identify two time scales.  The invasion time $t_{inv}$ (full red line), which scales like  $1/\sqrt{k_B T}$ at high temperatures, is the time required for $A$ particles to reach the right side of the container when the membrane is removed.  $\tau_d$ is the time required for the semi-permeable membrane to distinguish between particles, and is assumed to be temperature-independent (dashed lines). At $T_d$, $\tau_d = t_{inv}$. Well below this temperature, invasion is so slow that the membrane discriminates between $A$ and $B$ way before the mixing, hence the relevant thermodynamic properties are those of the true equilibrium, i.e., of a system with non-identical particles. Above $T_d$ the system reaches a long-lasting state before it can resolve between $A$ and $B$ and the thermodynamics during this period  is that of identical particles. As the similarity between $A$ and $B$ increases, so does $\tau_d$, so the corresponding $T_d$ decreases - in this cartoon $T_{d,1}$ corresponds to almost identical particles, $T_{d,3}$ to very different particles and $T_{d,2}$ is somewhere in between.  \label{fig2}}
\end{figure}

Accordingly, in this system the role played by $\tau_d$ depends on the temperature $T$. For a given $\delta r$ and $\tau_d$, if $T$ is too low the rate-determining step of $t_{eq}$ is $t_{inv}$ and the filtering by the semi-permeable membrane has only a minor effect on the dynamics. In other words, the system resolves the differences between $A$ and $B$ particles way before it equilibrates. When $T$ is high $t_{inv}$ shortens and $\tau_d$ dominates $t_{eq}$, as $\tau_d$ diverges so does the difference between $t_{eq}$ and $t_{peq}$. For each  $\tau_d$ there exist a characteristic crossover temperature, $T_d$. When $T \gg T_d$ there is a long period in which the system is close to its pseudo-equilibrium state, and during this period its thermodynamic behavior will be that of a system with identical particles. When $T \ll T_d$ there is no such period. These characteristics are illustrated in Figure~\ref{fig2}.

Other systems may have other factors that govern  $t_{eq} - t_{peq}$, but in general one expect this time scale to be shorter as the rate of stochastic or chaotic transitions increases, so it will decrease as the temperature of the system increases. Quantum mechanics suggests a general lower bound: since $T \Delta S$ is also an energy scale, the time for detection of one bit of entropy changes must be larger than $\hbar/k_B T$, the universally dissipative timescale considered in~\cite{hartnoll2015theory}. Therefore one may expect the existence of a temperature scale $T_d$, and the quantitative distinction between $T<T_d$ and $T>T_d$, to be a generic feature of thermal systems.

\subsection{Practical implications}

As explained, the relevant distinguishability time is system-specific or detector-specific. In any system $\tau_d$ is the time over which the physical interactions in \emph{that} system discriminate between particles, and it may be longer than its value in the presence of ideal or optimal detectors. Back to Figure~\ref{fig1}, if the holes in the membrane are of radius $r_A + \epsilon$ when $\epsilon$ is very small, the filtering time diverges even if $r_B \gg r_A$. Similarly, if the membrane is permeable for both $A$ and $B$ particles, the relevant distinguishability time is considered infinite. In both cases $T_d \to 0$.

In colloid systems, like those discussed by \citet{frenkel2014colloidal}, the forces that govern physics have typically nothing to do with the microscopic differences between colloids. Therefore the effective distinguishability time is much longer than any relevant observation time, $T_d$ is way below each relevant temperature and what appears to be the equilibrium properties of the system are in fact its pseudo-equilibrium properties; in the relevant statistical physics calculations one must introduce the Gibbs factor.

If the distinguishability time is the rate-limiting factor but is still much smaller than the observation time window, one expects a gradual process in which the system first relaxes to a quasi-steady-state that corresponds to identical particles, after a while it resolves the differences between them, entropy raises and the system equilibrates again (in the example of figure \ref{fig1}, a few $A$ particles cross the membrane, the right chamber equilibrates, then a few other particles cross and so on). Another non-trivial possibility is the case in which the thermalization process of colloids or other particles is relatively slow (e.g., when their dynamics is glassy) and is comparable with $\tau_d$, so $T \sim T_d$. In such a case the effective counting of microscopic permutations changes gradually during the equilibration process.  Glassy behavior in systems of non-identical particles, like those considered by~\cite{osmanovic2016neighborhood} may exhibit this feature.

\section{Discussion}

Both thermal physics and quantum physics distinguish between identical and non-identical particles. This is a binary distinction, independent of the actual physical differences between particles. Accordingly, as $\delta {\cal O}$ varies one expects an abrupt and discontinuous shift of the physical properties of a system at $\delta {\cal O} =0$, without any warning signal. Such a catastrophe does not make sense. When Gibbs' paradox is interpreted in that way (as in~\cite{van1984gibbs,denbigh1989gibbs,jaynes1992gibbs,allahverdyan2006explanation,dieks2011gibbs}), we cannot invoke the symmetries of the quantum wave-function to resolve it, as such a ''resolution" is just a regress of the same problem to a more fundamental level.

\citet{denbigh1989gibbs} have considered Gibbs' paradox and suggested a solution that does not rely on the properties of the microscopic constituents of the system. To do that, these authors analyzed the dynamic of a specific process, a reversible distillation of two ideal fluids that have different volatilities. To be consistent with the above notations, let us denote the difference between the vapor pressure of the two liquids by $\delta {\cal O}$.

Distillation takes place in a series of discrete steps, in each step the vapors (that are richer in the more volatile ingredient) are collected and then condensed back into a liquid (in their example, through an increase of pressure). By iterating this process one obtains two mixtures, one is rich in the less volatile liquid and one which is rich with the more volatile liquid. The decrease in entropy, $\Delta S$, during this reversible process was calculated by \citet{denbigh1989gibbs}. Importantly, when the number of steps $n$ goes to infinity the two liquids separate completely for any finite $\delta {\cal O}$. Therefore  $\Delta S(n \to \infty)$ (that depends only on the initial and on the final state) is independent of $\delta {\cal O}$ for any finite $\delta {\cal O}$, and jumps discontinuously to zero at $\delta {\cal O}=0$.

As explained by \citet{denbigh1989gibbs}, this catastrophe is a manifestation of the non-uniform convergence of $\Delta S(n)$ to its asymptotic value at $n \to \infty$. For any finite $n$, $\Delta S(n)$ is a continuous function of $\delta {\cal O}$ (like $x^n$ for $0 \le x \le 1$), but it converges to its asymptotic value non-uniformly (similarly, in the limit $n \to \infty$ the function $x^n \to 0$ for any $0 \le x < 1$, and jumps abruptly to one at $x=1$).

To elucidate the relationships between the measurement time problem considered here (Section \ref{sec3}) and techniques like distillation or enrichment, let us return to the setup suggested in Figure \ref{fig3}. Charged particles enter the system through a slit of width $\Delta y$, so the uncertainty in their initial $y$-momentum is $\hbar/\Delta y$. This uncertainty is independent of the charge of a particle, and we can think about the initial wavefunction as a zero-mean Gaussian in the momentum space $P_y$, whose width is that uncertainty. The electric force inside the capacitor transfers momentum to the particle, so the mean of the Gaussian increases but its width remains more or less constant. The separation between particles with different charges is good enough when the overlap between the corresponding wavefunctions $\psi(P_y^1)$ and $\psi(P_y^2)$ is negligible at the point where the particles exit the capacitor. To achieve that, the difference between the impulses felt by the particles must be larger than the initial uncertainty, and this requirement imposes a lower bound on the measurement time.

When the measurement time is too short, there is an alternative way to separate particles with different charges. By collecting those particles that have relatively high $P_y$ at the exit, one obtains an enriched mixture. Iterating this process again and again the separation becomes closer and closer to perfection. The measurement time and the number of iterations required for separation are thus the two sides of the same coin.

The analysis suggested through this paper is focused on continuous-time dynamics, does not distinguish between reversible and irreversible processes and treats the discontinuity problem in quantum and thermodynamic processes on equal footing. Yet the outcome is quite similar to the one obtained in \cite{denbigh1989gibbs}: the $\delta {\cal O}$ independent outcomes appear only in the asymptotic (infinite time) limit, and for any finite time physical observable has to be continuous functions of $\delta {\cal O}$. Therefore, although the identity of a pair of particles is a binary property, nature does not make jumps: as particles become more and more similar, the time required to distinguish between them diverges.

Had the world been classical, there would have been no lower bound on $\tau_d$. In such a world, one may implement a semi-permeable membrane that selects for any $\delta r$ at zero time. In a system equipped with these optimal measuring instruments $T_d$ will be infinite for distinguishable particles and will jump abruptly to zero when the particles become indistinguishable. Still, in a classical system that implements non-ideal detectors  (like the holes in the membrane discussed above) $T_d$ is finite for any $\delta {\cal O} \neq 0$ and the double-counting of permeations for slightly different particles must be avoided as long as $T<T_d$.

 In many physical scenarios there is a separation of timescales, equilibration time is either much shorter or much longer than $\tau_d$, so one either counts permutations as different microscopic states ($T<T_d$) or as a single state ($T>T_d$). When $\tau_d$ is close to the equilibration time, or when the observation time is much larger than $\tau_d$, particles that appear to be identical reveal their differences through the equilibration process. This may yield some new and interesting physics, yet to be explored.

\vspace{1.0cm}

\noindent \textbf{Acknowledgments}:  The author gratefully acknowledges fruitful discussions with David Kessler, Tamar Shnerb and Yitzhak Rabin.

\clearpage

\bibliography{gibbs_ref}

\clearpage

\appendix

\section{Detecting mass differences} \label{apa}

In the main text (the discussion surrounding Fig. \ref{fig3}) we considered the ability of a physical system to detect charge differences, e.g., to distinguish between electric charge $q$ and another charge $q+\delta q$. Basically, we assumed the existence of a field (in that case, electric field ${\cal E}$) that couples to the corresponding charge, so in a fixed field the forces on different charges ($q$ and $q+\delta q$) differ by $\delta q {\cal E}$. When the force is applied through a period $\Delta t$, its contributions to the momenta of the two particles (in the direction of the applied force, here chosen to be the $y$-direction) differ by $\Delta P_{\cal E} = \delta q {\cal E} \Delta t$. To resolve between the two charges $\Delta P_{\cal E}$ must be larger than the quantum mechanical uncertainty $\Delta P_{qm} = \hbar/\Delta y$, where $\Delta y$ is the width of the entrance slit. Therefore, $\Delta t$ must be greater from, or equal to, $\hbar/\delta q {\cal E} \Delta y$.

These considerations suggest that and charge difference $\delta q$ may be resolved in an arbitrary short time $\delta t$, provided that $\Delta y$ is large enough, $\Delta y > \hbar/\delta q {\cal E} \delta t$.  The restriction $\Delta t \ge \hbar/\Delta E$ is related to energy, not to charge. In the deflection experiment the energy difference is $\delta q {\cal E} \Delta y$, so it grows linearly with $\Delta y$, and by taking the slit width to infinity one may reduce the measurement time to zero.

The same considerations apply to any other physical property or charge, with the exception of \emph{mass}.

Let us try to distinguish between two particles, one of mass  $m$ and the other of mass $m+\delta m$, using the deflection setups of Figure \ref{fig3}, where the field is a fixed gravitational field $g$ pointing in the $y$ direction. The above discussion appears to suggest that one may resolve between $m$ and $m+\delta m$ as long as,
\begin{equation} \label{eq7}
\Delta t_{qm} \ge \frac{\hbar}{g \, \delta m \, \Delta y}
\end{equation}
so the measurement time, again, goes to zero when $\Delta y$ diverges.

This conclusion must be wrong, since a mass scale is inherently related to an energy scale via $\Delta E = \delta m \, c^2$. On the contrary, Eq. (\ref{eq7}) implies that  by increasing $\Delta y$ one may break the limitation imposed by the energy-time uncertainty relationships.

To address this question the principles of general relativity must be invoked. The lapse of time changes along the gradient of the gravitational potential. When a weak gravitational field $g$ is applied in the $y$ direction,
\begin{equation}
\Delta t(y) = \left( 1+ \frac{g y}{c^2}\right) \Delta t(y=0).
\end{equation}

Therefore, the uncertainty in $\Delta y$ is translated to an uncertainty in the duration of the experiment. Without the gravitational effect on time, the minimum duration of the measurement, $\Delta t_{qm}$, is dictated by Eq. (\ref{eq7}). If this is the measurement time for a particle that enters through the lower part of the slit, at $y=0$, then the measurement time for a particle that enters in the higher part of the slit will be,
\begin{equation}
\Delta t_{qm+g}  = \frac{g \Delta y}{c^2} \frac{\hbar}{g \, \delta m \, \Delta y} = \frac{\hbar}{\delta m \, c^2},
\end{equation}
as requested.

\end{document}